\documentclass[%
 reprint,
 amsmath,amssymb,
 aps,
]{revtex4-1}

\usepackage{graphicx}
\usepackage{dcolumn}
\usepackage{bm}

\newcommand{\ket}[1]{\ensuremath{\left|#1\right\rangle}}

\begin{document}

\preprint{APS/123-QED}

\title{Efficient Quantum Digital Signatures without Symmetrization Step}

\author{Yu-Shuo Lu}
\author{Xiao-Yu Cao}
\author{Chen-Xun Weng}
\author{Jie Gu}
\author{Yuan-Mei Xie}
\author{Min-Gang Zhou}
\author{Hua-Lei Yin}\email{hlyin@nju.edu.cn}
\author{Zeng-Bing Chen}\email{zbchen@nju.edu.cn}
\affiliation{National Laboratory of Solid State Microstructures, School of Physics and Collaborative Innovation Center of Advanced Microstructures, Nanjing University, Nanjing 210093, China.}

\begin{abstract}
Quantum digital signatures (QDS) exploit quantum laws to guarantee non-repudiation, unforgeability and transferability of messages with information-theoretic security. Current QDS protocols face two major restrictions, including the requirement of the symmetrization step with additional secure classical channels and quadratic scaling of the signature rate with the probability of detection events.
 Here, we present an efficient QDS protocol to overcome these issues by utilizing the classical post-processing operation called post-matching method. Our protocol does not need the symmetrization step, and the signature rate scales linearly with the probability of detection events. Simulation results show that the signature rate is three orders of magnitude higher than the original protocol in a 100-km-long fiber.
This protocol is compatible with existing quantum communication infrastructure, therefore we anticipate that it will play a significant role in providing digital signatures with unconditional security.
\end{abstract}

\maketitle
\section{Introduction}\label{sec1}
Cryptography is essential for uncounted amount of applications that rely on non-repudiation, integrity and confidentiality of data.
The two pillars of modern cryptography are encryption and digital signatures~\cite{diffie1976new}, where encryption guarantees confidentiality and digital signatures provide integrity and non-repudiation.
Traditionally, public-key cryptography algorithms, such as the Rivest-Shamir-Adleman algorithm~\cite{rivest1978method}, are designed to simultaneously provide encryption and digital signature service. However, relying on the computational difficulty of certain mathematical problems, public-key cryptosystems are usually vulnerable to quantum computing attacks~\cite{shor1994algorithms}.
Quantum key distribution (QKD) allows two remote users to share a secret key string with information-theoretic security~\cite{bennett1984proceedings,ekert1991quantum}. By combining one-time pad encryption~\cite{shannon1949communication} and QKD, one can implement information communication with perfect confidentiality~\cite{chen2009field}. In addition, the direct transmission of private information is made possible in principle by quantum encryption~\cite{qi2019implementation}.

Digital signatures are widely applied in e-mails, electronic commerce and software distribution to ensure data integrity and non-repudiation~\cite{Pirandola:20}. Similar to QKD used for encryption service, quantum digital signatures (QDS) are expected to provide information-theoretic security to sign documents. The first QDS protocol was proposed in 2001~\cite{gottesman2001quantum}, but it is unfeasible because of challenging experimental requirements. In the next decade or so, great efforts have been made in developing QDS protocols and an important achievement was the removal of the requirement of quantum memory~\cite{clarke2012experimental,dunjko2014quantum, collins2014realization,wallden2015quantum,Croal:2016:Free}. Nevertheless, the security analysis of the early QDS protocols are  based on secure quantum channels, i.e., there is no eavesdropping, which is a conflicting assumption.
In 2016, two independent QDS protocols were proposed and proved to be secure against the general attacks without the assumption of secure quantum channels~\cite{yin2016practical,PhysRevA.93.032325}. Importantly, their experimental devices and techniques have already been widely employed in QKD.
These two protocols are important steps towards practical QDS~\cite{Pirandola:20}. The one in~\cite{yin2016practical} is based on non-orthogonal encoding. The other~\cite{PhysRevA.93.032325} utilizes orthogonal encoding, which results in the need of an additional symmetrization step in the protocol. In addition, great achievements have been made in the experimental and theoretical research of information-theoretically secure QDS  ~\cite{puthoor2016measurement,yin2017experimental102,collins2017experimental,yin2017experimental,roberts2017experimental,zhang2018proof,Thornton:2019:CV,an2019practical,ding2020280,zhang2020twin,wang2015security,wang2017postprocessing}, including the field test of measurement-device-independent (MDI) QDS~\cite{yin2017experimental}.

For the orthogonal encoding based protocol~\cite{PhysRevA.93.032325} (see also~\cite{puthoor2016measurement,zhang2020twin}), the need of the symmetrization step, which requires an additional secure classical channel, is the main issue. Currently, secure classical channels can only be realized by combining QKD and one-time pad encryption. The symmetrization step will consume $6L$ bits of secret key generated by QKD if one uses $L$ bits as the signature~\cite{yin2017experimental}. Specially, in the worst case where the signer is located in the middle of two receivers, the low secret key rate of QKD between the two receivers severely limits the real-time signature rate of QDS.
Besides, considering a quantum network with $J$ users, there will be a need of $J(J-1)/2$ secure classical channels~\cite{roberts2017experimental}, which is an unrealistically high amount in a real quantum network. For the non-orthogonal encoding based protocol~\cite{yin2016practical}, it does not require the symmetrization step. However, the signer has to send the same quantum states to two receivers. Only coincidence detection events, i.e., the two receivers both have click, are valid events.
Let $\eta$ be the probability that one receiver has click, if the signer sends $N$ quantum states to the two receivers, there will be only $\eta^2 N$ valid events. Therefore, the signature rate quadratically scales with the probability of detection events.

Here, inspired by the original protocol in~\cite{yin2016practical}, we propose an efficient quantum digital signature protocol without symmetrization step. A novel classical post-processing operation called post-matching method is exploited in our protocol.
With the help of the post-matching method, the requirement of coincidence detection is removed. Given that the signer sends $N$ quantum states to receivers, there will be $\eta N$ valid events.
Therefore the signature rate decays linearly with the probability of detection events. Simulation results show that the signature rate of our scheme is $2$ or even $3$ orders of magnitude higher than that of Ref.~\cite{yin2016practical} in large attenuation case, and is comparable to orthogonal encoding based protocol.

\section{Protocol Description}\label{sec2}
There are three participants in our protocol, namely the signer Alice and the receiver Bob and Charlie. As determined by Alice, either Bob or Charlie can be the authenticator of the signature, and the other becomes the verifier. There are noisy insecure quantum channels connecting Alice-Bob and Alice-Charlie, and authenticated classical channels between the three participants.
There are three stages in our protocol: key generation, estimation and messaging.  In our protocol, the three stages can be performed separately, which means they can generate raw keys and store them for a long time, and continue the estimation and messaging stage whenever Alice wants to sign the message. This makes our protocol more practical.

Our protocol exploits non-orthogonal encoding to generate logical bits~\cite{scarani2004quantum}. There are four quantum states: $\ket{H}$, $ \ket{V}$, $\ket{+}$ and $\ket{-}$, where $\ket{H}$ and $ \ket{V}$ are the eigenstates of the Pauli Z operator and $\ket{+}$ and $\ket{-}$ ( $\ket{\pm}=\frac{1}{\sqrt{2}}($\ket{H}$\pm $ \ket{V}$)$) are the eigenstates of the Pauli X operator. These four quantum states can be arranged into four sets: $\{\ket{H}, \ket{+}\}$, $\{\ket{+}, \ket{V}\}$, $\{\ket{V}, \ket{-}\}$, $\{\ket{-}, \ket{H}\}$, where the first state in each set is encoded with bit value 0 and the second is encoded with 1. Alice randomly sends quantum states to receivers and assigns each quantum state to a set. The receivers randomly choose Z or X basis to perform polarization measurement on each quantum state. If the measurement outcome is orthogonal to one of the states in the set, the receiver obtains a conclusive result with bit value 0 or 1, otherwise the receiver obtains an inconclusive result, denoted by $\bot$. Note that the set assigned by Alice should contain the quantum state she sent. For example, if Alice sends $\ket{H}$, she should assign it to set $\{\ket{H}, \ket{+}\}$ or $\{\ket{-}, \ket{H}\}$. When she assigns it to the set $\{\ket{H}, \ket{+}\}$ and Bob's measurement outcome is $ \ket{-} $ ($\ket{V}$), Bob obtains a conclusive result with bit value $0$ $(1)$.

\begin{figure}  [t]
	\centering
	\includegraphics[width=8cm]{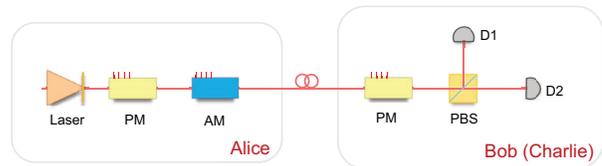}
	\caption{\label{fig1}Schematic diagram of a setup for our QDS protocol.  Alice randomly prepares one of the four Bennett-Brassard 1984~\cite{bennett1984proceedings} (BB84) states with phase-randomized weak coherent-state source  and sends them to Bob (Charlie). Bob (Charlie) performs polarization measurement in the $Z$ or $X$ basis.   PM: polarization modulator;
		AM: amplitude modulator;
		PBS: polarization beam splitter;
		D1-D2: single photon detectors;}
	\label{{fig1}}
\end{figure}

The decoy-state method~\cite{wang2005beating,lo2005decoy} with three intensities is exploited to deal with photon-number-splitting attack for coherent state source. Data from the decoy state and vacuum state will be used for parameter estimation, and only data from the signal state will be used as test bits and secret keys. The setup for our QDS protocol is presented in Fig.~\ref{fig1}. In the following part of the paper, we use superscripts $c$, $u$, $t$, $*$, overline (underline), to denote conclusive results, untest bits, test bits, expected value and the upper bound (lower bound) of expected value, respectively. We also use subscripts $P$ ($P\in\{A, B, C\}$), $11$ and $\lambda$ ($\lambda\in\{\mu, \nu, 0\}$) to denote Alice (Bob, Charlie), single-photon pair components and intensity respectively. Detailed descriptions of our protocol are given below.

~\noindent{\it{1.~Key generation.}}
(1) Alice randomly selects a quantum state $\{\ket{H}, \ket{V}, \ket{+}, \ket{-}\}$ with the same possibility and an intensity $\{\mu, \nu, 0\}$ (signal, decoy and vacuum state) with possibilities $p_{\mu}$, $p_{\nu}$ and $p_0 $ respectively.  For each possible message $m$ ($m=0$ or $1$), Alice prepares two different quantum state sequences with length $N$, namely $A_{B,m}$, and $A_{C,m}$.
Alice sends $A_{B,m}$ to Bob and $A_{C,m}$ to Charlie through insecure quantum channels.

(2) For each quantum state, Bob and Charlie randomly choose X or Z basis to perform polarization measurement.
Bob announces all the click events in $A_{B,m}$ through authenticated classical channel. Alice and Bob discard all the data that has no click. They keep the left data of length $n$, denote as $S_{AB,m}$ (kept by Alice) and $S_{B, m}$ (kept by Bob). Alice and Charlie perform the same step. As a result, Alice has four data strings $S_{AB,0}$, $S_{AB,1}$, $S_{AC, 0}$ and $S_{AC, 1}$, Bob (Charlie) has two strings $S_{B, 0}$($S_{C, 0}$) and $S_{B, 1}$($S_{C, 1}$).
Since Alice randomly and independently chooses quantum states, the quantum states that Bob and Charlie receive are uncorrelated.

(3) Alice announces the intensity information of all pulses. According to the intensity information, the three participants divide each of their data strings into three strings, namely $\mu$ string, $\nu$ string and $0$ string. For example, Bob divides $S_{B, m}$ into $S_{B, m}^\mu$, $S_{B, m}^\nu$ and $S_{B, m}^0$.

\begin{figure}[t]
	\centering
	\includegraphics[width=8cm]{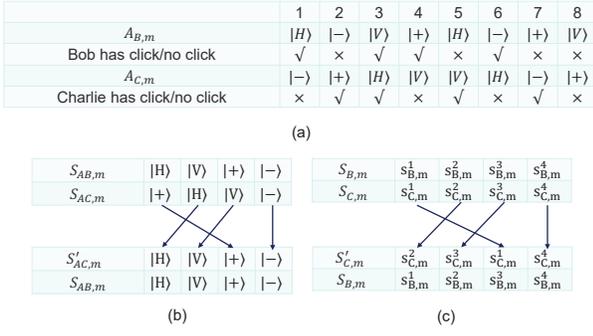}
	\caption{\label{fig2} Schematic diagram of the post-matching method. For simplicity, we temporarily omit the superscript $\lambda$.
	(a) Alice sends $A_{B,m}$ to Bob and sends $A_{C,m}$ to Charlie. Only part of quantum states can be detected due to channel loss and imperfect detection. We use `$\checkmark$' (`$\times$') to denote the detector has click (no click). They discard the data that has no click and keep the remaining data.
	(b) Alice changes the order of $S_{AC,m}$ into $S'_{AC,m}$. Alice informs Charlie about the procedure of changing the order of data. For example, if $S_{AB, m}$$=$ $\{s_{AB,m}^1$, $s_{AB,m}^2$, $s_{AB,m}^3$, $s_{AB,m}^4\}$$=$$\{\ket{H}$, $\ket{V}$, $\ket{+}$ $\ket{-}\}$, $S_{AC, m}$$=$$\{s_{AC,m}^1$, $s_{AC,m}^2$, $s_{AC,m}^3$, $s_{AC,m}^4\}$$=$$\{\ket{+}$, $\ket{H}$, $\ket{V}$, $\ket{-}\}$, Alice should change the order of elements in $S_{AC,m}$ into $\{s_{AC, m}^{2}$, $s_{AC, m}^{3}$, $s_{AC, m}^{1}$, $s_{AC, m}^{4}\}$. She also asks Charlie to change the order of elements in $S_{C, m}$ into $\{s_{C, m}^{2}$, $s_{C, m}^{3}$, $s_{C, m}^{1}$, $s_{C, m}^{4} \}$.
	(c) Charlie changes the order of $S_{C,m}$ as instructed by Alice. Note that $S_{C,m}$ is the measurement result of $S_{AC,m}$.
	Since the order of elements in $S_{C,m}$ and $S_{AC,m}$ are changed with the same procedure,  $S'_{C,m}$ is the measurement result of $S'_{AC,m}$. As a result, although Alice sends different quantum state sequences $A_{B,m}$ and $A_{C,m}$, after post-matching, it is equivalent to two identical sequences $S_{AB,m}$ and $S'_{AC,m}$ arrive at Bob and Charlie.
}
\end{figure}

(4) For the data strings corresponding to each intensity $\lambda$ ($\lambda\in \{\mu, \nu, 0\}$), Alice takes $S_{AB,m}^\lambda$ as the reference and changes the order of elements in $S_{AC,m}^\lambda$. Denote the changing result as $S'^\lambda_{AC,m}$, Alice should make $S^\lambda_{AB,m}$ and $S'^\lambda_{AC,m}$ identical. Without loss of generality, she requests Charlie to change the order of elements in $S_{C,m}^\lambda$ into the same order.
We call this the post-matching method. After post-matching, the data obtained by Bob and Charlie can be correlated. Detailed description of post-matching method is given in Fig.~\ref{fig2}.

(5) Using the rules for generating logical bits, Alice randomly assigns each element in $S_{AB, m}^\lambda$ a set. The three participants `translate' their data strings into raw key strings denoted as $K_{A, m}^\lambda$, $K_{B, m}^\lambda$ and $K_{C, m}^\lambda$. Note that they do not announce which bits are conclusive results.

~\noindent{\it{2.~Estimation.}}
(1) The signer Alice chooses the desired authenticator of the signature, and the other participant automatically becomes the verifier. Here we assume Bob is the authenticator. The three participants publicly announce all data of $\nu$ strings and $0$ strings and the value of $n_\lambda$ (the length of $\lambda$ string). They estimate bit error rate of single-photon pair components in $\mu$ strings using decoy state method. The verifier Charlie randomly selects a proportion of $t$ in the $\mu$ string as test bits and asks Alice to publicly announce the value of these bits.
We denote test bit strings as $K_{A,m}^{t}$, $K_{B,m}^{t}$ and $K_{C,m}^{t}$. Let $E_B^{ct}$ ($E_C^{ct}$) be the mismatch rate of conclusive results between $K_{A,m}^t$ and $K_{B,m}^t$ ($K_{A,m}^t$ and $K_{C,m}^t$).
Bob and Charlie calculate $E_B^{ct}$ and $E_C^{ct}$. Note that when $E_B^{ct}$ or $E_C^{ct}$ gets too high, the signing process of this round is highly possible to fail. In this case they abort the protocol. In addition, Bob and Charlie calculate the proportion of conclusive results in $K_{B,m}$ and $K_{C,m}$ , denoted as $P_B^c$ and $P_C^c$, respectively. If $P_B^c$ or $P_C^c$ shows a big deviation from the ideal value $\frac{1}{4}$, they will also abort the protocol. The three participants discard the test bits and keep the remaining untest bits in $\mu$ strings with length $n^u$. We denote these untest bit strings as $K_{A,m}^u$, $K_{B,m}^u$ and $K_{C,m}^u$. They will be used as secret keys to sign the message in the messaging stage.

(2) Bob and Charlie announce $\{E_B^{ct}$, $P_B^c\}$ and $\{E_C^{ct}$, $P_C^c\}$. The three participants publicly negotiate to determine the values of authentication security threshold $T_a$ and verification security threshold $T_v$.

~\noindent{\it{3.~Messaging.}}
(1) To sign a one-bit message $m$, Alice sends the message  and the corresponding secret key $\{m, K_{A,m}^u\}$ to the authenticator Bob. Bob calculates the mismatch rate of conclusive results between $K_{A,m}^u$ and $K_{B,m}^u$, which is denoted as $E_B^{cu}$. If $E_B^{cu}< T_a$, Bob accepts the message and forwards $\{m, K_{A,m}^u\}$ to Charlie, otherwise he rejects the message and announces to abort the protocol.

(2) After receiving $\{m, K_{A,m}^u\}$ forwarded by Bob, Charlie calculates the mismatch rate of conclusive results $E_C^{cu}$ between $K_{A,m}^u$ and $K_{C,m}^u$. Charlie accepts the message if $E_C^{cu}<T_v $. When both Bob and Charlie accept the message, Alice successfully signs the message.

\section{Security Analysis}

In our protocol, Alice randomly chooses unrelated quantum states to send to Bob and Charlie.
After post-matching, it is equivalent to Alice simultaneously sending the same quantum states to Bob and Charlie.
To perform post-matching, Alice exposes information of order, but does not leak information of quantum states.
In this case, eavesdroppers can not obtain more information of quantum states compared with the case where Alice actually sends two copies of quantum states. Thus the security analysis of our protocol can directly follow the lines in Ref.~\cite{yin2016practical}.
In the three-participant scenario, transferability and nonrepudiation are equivalent. Accordingly, there are three security criteria: robustness, security against forging and security against repudiation. For simplicity, we just briefly present our results.
For more detail, refer to Ref.~\cite{yin2016practical}.

~\noindent{\it{1. Robustness.}}
The robustness means the probability of an honest abort $\epsilon_{rob}$. In messaging stage, Bob rejects the message sent by Alice when $E_B^{cu}>T_a$. In the case of finite sample size, the robustness can be quantified by exploiting random sampling without replacement theorem~\cite{yin2020tight}.

~\noindent{\it{2. Security against forging.}}
In a forgery attack,  Bob sends the message he wishes to forge and its corresponding secret key $\{m, K_{BF,m}\}$ to Charlie. The forgery attack is successful if Charlie accepts Bob's forged message. An honest Bob knows only about $\frac{1}{4}$ of conclusive results in $K_{A,m}^u$. If Bob is an adversary, his optimal strategy is to acquire information of quantum states Charlie receives as much as possible, which is equivalent to the eavesdropping attack of Eve in four-state Scarani-Acin-Ribordy-Gisin 2004 (SARG04) QKD with two-photon source~\cite{scarani2004quantum, tamaki2006unconditionally,yin2016security}.

We assume only single-photon pairs that Alice sends to Bob and Charlie are secure. In this case, Bob and Charlie both receive a single-photon. Using Chernoff bound~\cite{chernoff1952measure}, the probability of a successful forgery attack $\varepsilon_{for}$ can be given by
\begin{equation}
	\varepsilon_{for}=\exp\left[ -\frac{(E_{BF11}^*-T_{v11})^2}{2E_{BF11}^*} n^{cu}_{11}\right],
\end{equation}
where $E_{BF11}^*$ is the expected value of minimum mismatch rate of conclusive results of the single-photon pair components between $K_{A,m}^u$ and $K_{BF,m}^u$, $T_{v11}=T_vn^{cu}/n^{cu}_{11}$ is the error rate threshold of single-photon pair, $n^{cu}=(1-t)n_\mu^c$ is the number of conclusive results in $K_{C,m}^u$, $n^{cu}_{11}=(1-t)s_{C11}^{c\mu}$ is the number of single-photon pair components in $K_{C,m}^u$ and $s_{C11}^{c\mu}$ is the number of events in which Bob and Charlie both receive a single-photon in $\mu$ string and Charlie has a conclusive result.

To obtain the value of $E_{BF11}^*$, one should exploit decoy state method to estimate $s_{C11}^{c\mu}$ and  $t_{C11}^{c\mu}$, where $t_{C11}^{c\mu}$ is the number of events that Bob and Charlie both receive a single-photon in $\mu$ string, Charlie has a conclusive result, and his classical bit is mismatching with Alice's. We use $n_{P\lambda}$ to denote the number of detection events of the participant $P$ of intensity $\lambda$ and $m_{P\alpha}$ to denote the number of mismatching bits. The expected value of parameter $x$ can be acquired by the variant of Chernoff bound~\cite{yin2020tight}: $\overline{x}^*=x+\beta+\sqrt{2\beta x +\beta^2}$ and $\underline{x}^*=x-\frac{\beta}{2}-\sqrt{2\beta x +\frac{\beta^2}{4} }$ with $\beta=\ln\frac{1}{\varepsilon_1}$, where $\epsilon_1$ is the failure probability of the Chernoff bound.

Separately consider the process that Alice sends pulses to Bob and to Charlie, we have
\begin{equation}\label{B1}
	s_{C1}^{c\mu^*}\ge \frac{p_{\mu}  e^{-\mu}}{\nu(\mu-\nu)}\left[\mu^2e^{\nu} \frac{\underline{n}_{C\nu}^{c^*}}{p_{\nu}}-\nu^2 e^{\mu} \frac{\overline{n}_{C{\mu}}^{c^*}}{p_{\mu}}+
	(\nu^2-\mu^2)\frac{\overline{n}_{C0}^{c^*}}{p_{0}} \right],
\end{equation}
and
\begin{equation}\label{B2}
	s_{B1}^{\mu^*}\ge
	\frac{p_{\mu} e^{-\mu}}{\nu(\mu-\nu)}\left[\mu^2e^{\nu} \frac{\underline{n}_{B\nu}^*}{p_{\nu}}-\nu^2 e^{\mu} \frac{\overline{n}_{B\mu}^*}{p_{\mu}}+
	(\nu^2-\mu^2)\frac{\overline{n}_{B0}^*}{p_{0}} \right],
\end{equation}
where $s_{B1}^{\mu}$  is the number of single-photon events in Bob's $\mu$ string and $s_{C1}^{c\mu}$ is is the number of conclusive single-photon events in Charlie's $\mu$ string. $s_{C11}^{c\mu^*}$ can be given by
\begin{equation}
	s_{C11}^{c\mu^*}\ge \underline{s}_{C1}^{c\mu^*} \times
	\frac{\underline{s}_{B1}^{\mu^*} }{\overline{n}_{B\mu}^*}.
\end{equation}
Bring in Eqs.(\ref{B1}) (\ref{B2}), we have
\begin{align}
	s_{C11}^{c\mu^*}&\ge
\frac{p_{\mu}^2  e^{-2\mu}}{\nu^2(\mu-\nu)^2\overline{n}_{B\mu}^*}	
\left[\mu^2e^{\nu} \frac{\underline{n}_{C\nu}^{c^*}}{p_{\nu}}-\nu^2 e^{\mu} \frac{\overline{n}_{C{\mu}}^{c^*}}{p_{\mu}}+(\nu^2-\nonumber\right.\\
&\left.\mu^2)\frac{\overline{n}_{C0}^{c^*}}{p_{0}} \right]\times\left[\mu^2e^{\nu} \frac{\underline{n}_{B\nu}^*}{p_{\nu}}-\nu^2 e^{\mu} \frac{\overline{n}_{B\mu}^*}{p_{\mu}}+
(\nu^2-\mu^2)\frac{\overline{n}_{B0}^*}{p_{0}} \right].	
\end{align}
We also have
\begin{equation}\label{B4}
	t_{C1}^{c\mu^*}\le \frac{p_{\mu}\mu e^{-\mu}}{\nu}(e^{\nu} \frac{\overline{m}_{C\nu}^{c^*}}{p_{\nu}}
	-\frac{\underline{n}_{C0}^{c^*}}{2p_{0}}  ),
\end{equation}
and
\begin{equation}\label{B5}
	s_{B1}^{\mu^*}\le \frac{p_{\mu}\mu e^{-\mu}}{\nu}(e^{\nu} \frac{\overline{n}_{B\nu}^*}{p_{\nu}}
	-\frac{\underline{n}_{B0}^*}{p_{0}}),
\end{equation}
where $t_{C1}^{c\mu}$ is the number of single-photon errors of Charlie's conclusive results in $\mu$ string with respect to Alice. $t_{C11}^{c\mu^*}$ can be given by:
\begin{equation}
	t_{C11}^{c\mu^*}\le \overline{t}_{C1}^{c\mu^*} \times \frac{\overline{s}_{B1}^{\mu^*} }{\underline{n}_{B\mu}^*}.
\end{equation}
Bring in Eqs. (\ref{B4}) (\ref{B5}), we have
\begin{align}
	t_{C11}^{c\mu^*}\le \frac{p_{\mu}^2\mu^2 e^{-2\mu}}{\nu^2\underline{n}_{B\mu}^*}\left(e^{\nu} \frac{\overline{m}_{C\nu}^{c^*}}{p_{\nu}}
-\frac{\underline{n}_{C0}^{c^*}}{2p_{0}}   \right)\times
\left(e^{\nu} \frac{\overline{n}_{B\nu}^*}{p_{\nu}}
-\frac{\underline{n}_{B0}^*}{p_{0}}\right).
\end{align}
~\noindent{\it{3. Security against repudiation.}}
\label{app_rep}
Alice successfully repudiates the message when Bob accepts the message while Charlie rejects it, i.e., $E_B^{cu}<T_a $ and $E_C^{cu}>T_v$. Alice does not know which bits are conclusive results for Bob (Charlie) and has to treat each bit in $K_{B,m}$ and $K_{C,m}$ with the same status.
For Bob and Charlie, the difference between $E_B^{cu}$ and $E_C^{cu}$ can be restricted by inequalities of the relative Hamming distance. The upper bound of the relative Hamming distance between $K_{B,m}^{cu}$ and $K_{C,m}^{cu} $ (denoted by $\overline{\Delta}_{BC}^{cu}$) can be given by using the random sampling without replacement theorem~\cite{yin2020tight}.
The probability of successful repudiation $\varepsilon_{rep}$ can be given by
\begin{equation}
	\varepsilon_{rep}=\exp\left[-\frac{\left(A-P_B^cT_a\right)^2}{2A}n^u\right],
\end{equation}
where $A$ is the solution of the following equation and inequalities:
\begin{equation}
	 \frac{\left[P_C^cT_v-P_C^c\left(\frac{\overline{\Delta}_{BC}^{cu}}{n^{cu}}+\frac{A}{P_B^c}\right)\right]^2}{3P_C^c\left(\frac{\overline{\Delta}_{BC}^{cu}}{n^{cu}}+\frac{A}{P_B^c}\right)}=\frac{(A-P_B^cT_a)^2}{2A},
\end{equation}
with $ P_B^cT_a<A<P_B^c\left(T_v-\frac{\overline{\Delta}_{BC}^{cu}}{n^{cu}}\right)$.

The overall secrecy is:
\begin{equation}
	\varepsilon_{tot}=11\epsilon_1+\epsilon_2 +\varepsilon_{for}+\varepsilon_{rob}+\varepsilon_{rep},
\end{equation}
where $\epsilon_2$ is the  failure probability of random sampling without replacement.

\begin{figure}
	\centering
	\includegraphics[width=8cm]{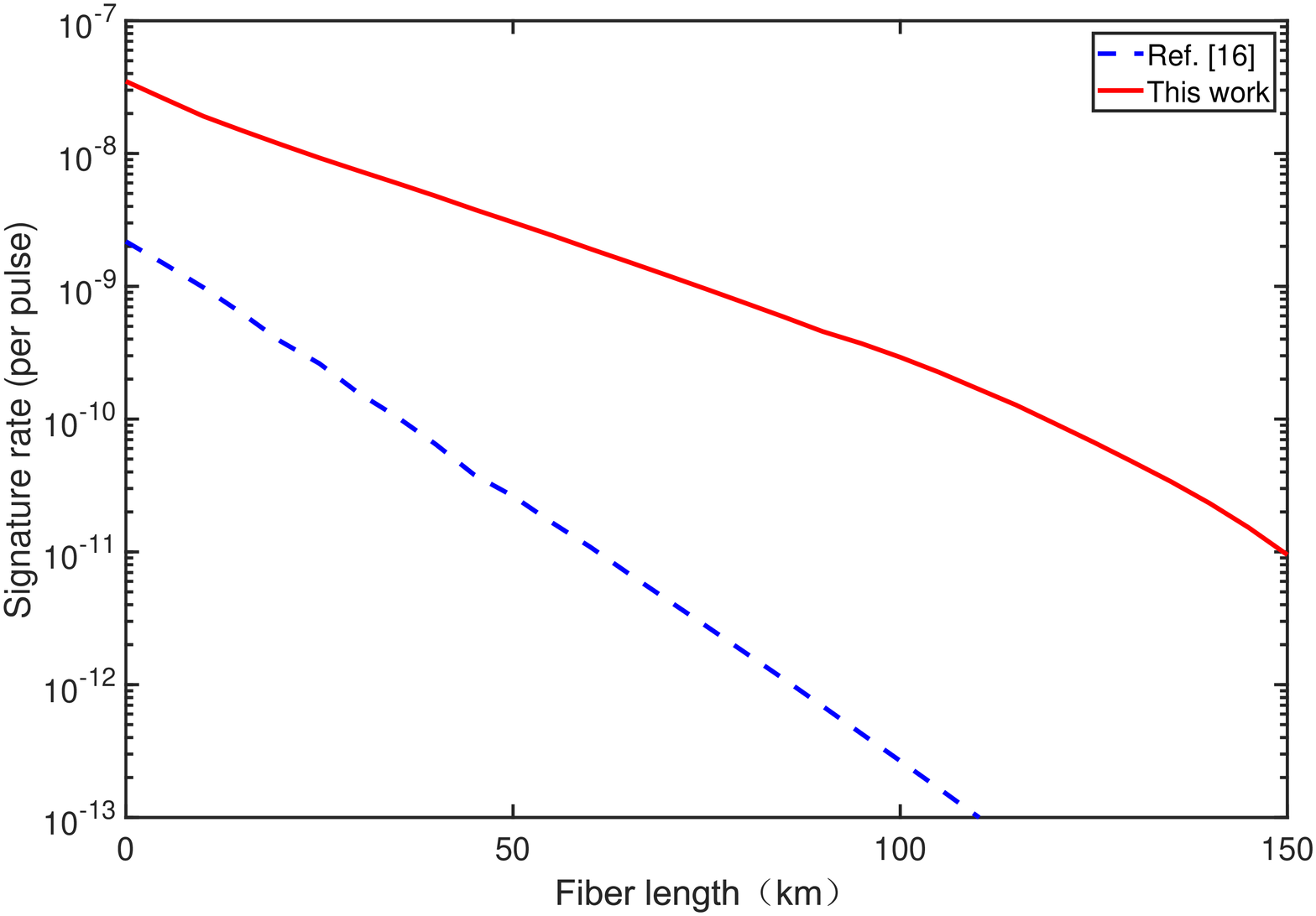}
	\caption{\label{fig3} Simulation of our QDS protocol. Numerically optimized signature rates are presented in logarithmic scale.  The detection efficiency is $52\%$, the dark counting rate is $1.3 \times 10^{-7}$, the basis misalignment rate is $0.15\%$, the insert loss is $1.2$ dB, and the loss coefficient of fiber is $0.194$ dB/km.}
\end{figure}
\section{Performance}\label{sec3}

In order to show the performance of our protocol, we simulate a fiber-based QDS system. Define signature rate $R:=\frac{1}{2N}$, where $2N$ is the minimum number of pulses required to securely sign a one-bit message. Fig.~\ref{fig3} shows the signature rate $R$ as a function of transmission distance. We consider the case where channels between Alice-Bob and Alice-Charlie are symmetric.

The security bounds are set to $\varepsilon_{for} \le 10^{-10}$,  $\varepsilon_{rob}\le 10^{-10}$,  $\varepsilon_{rep}\le 10^{-10}$ and $\varepsilon_1=\varepsilon_2\le (10^{-9}-3\times 10^{-10})/12$. We numerically optimize the minimum number of pulses required to securely sign a one-bit message with the free parameters $\{\mu, \nu, p_{\mu}, p_{\nu}, t\}$ by global search algorithm. For a fair comparison, we simulate the performance of the original protocol in Ref.~\cite{yin2016practical} with the same experimental parameters. As shown in Fig.~\ref{fig3}, the solid red line represents the signature rate of this work and the blue dashed line represents the original QDS protocol. Obviously, our protocol requires far less number of pulses to sign a one-bit message.
Specifically, at $50$ km and $100$ km, our protocol requires $3.3 \times 10^8 $ and $3.4 \times 10^{9}$ pulses to sign a one-bit message, but the protocol in Ref.~\cite{yin2016practical} requires $3.8 \times 10^{10}$ and $ 3.7 \times 10^{12}$ pulses. The signature rate of our protocol is 2 or even 3 orders of magnitude higher than the original protocol at long distance.


\begin{figure}[t]
	\centering
	\includegraphics[width=8cm]{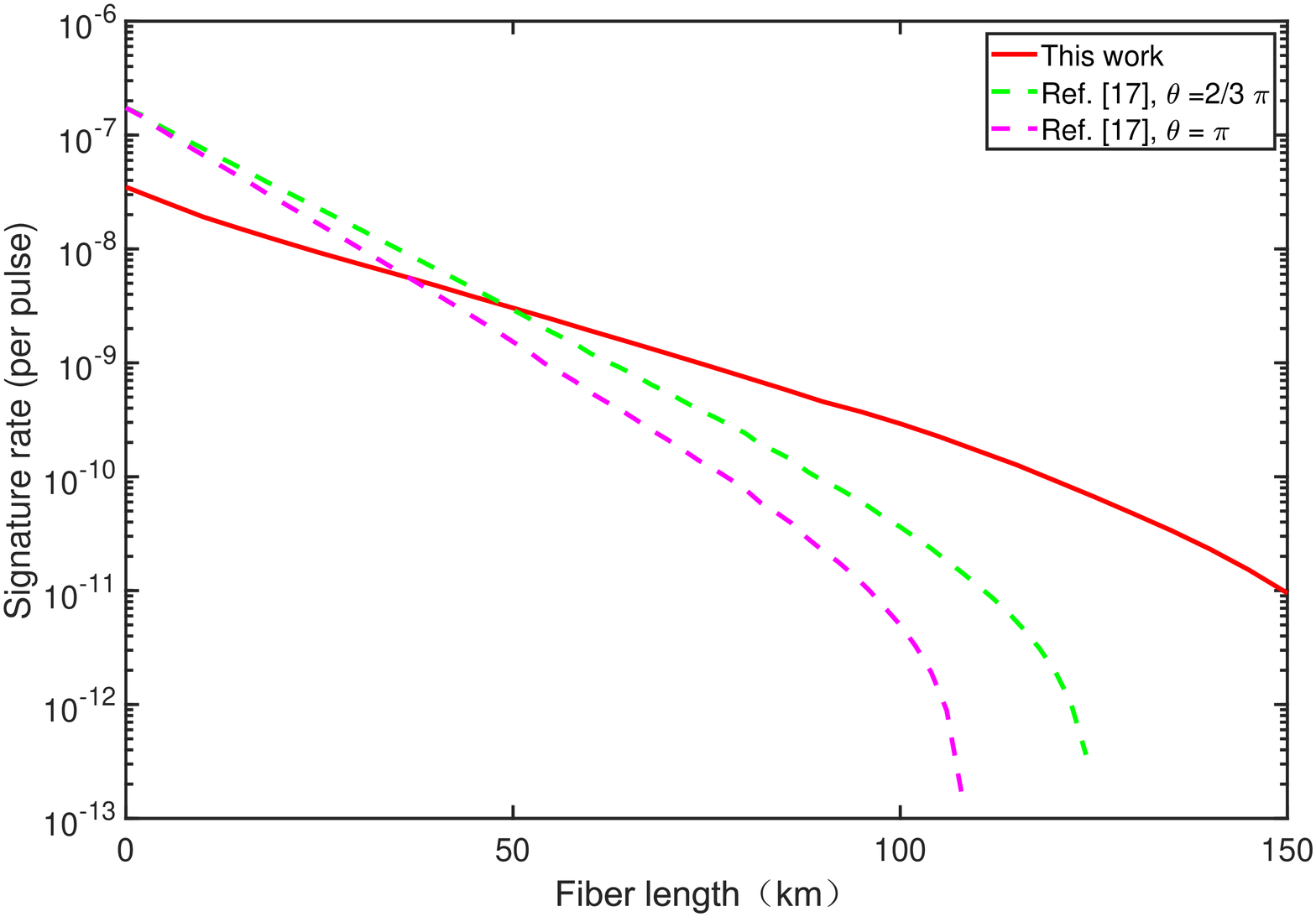}
	\caption{\label{fig4} Comparison of our QDS protocol and orthogonal encoding based protocol \cite{PhysRevA.93.032325}. The security bounds and experiment parameters are the same as Fig. \ref{fig3}. }
\end{figure}

We also simulate the performance of orthogonal encoding based protocol  \cite{PhysRevA.93.032325} with the cost of symmetrization taken into consideration. Assume Bob and Charlie utilize three-intensity decoy-state BB84 QKD protocol to perform symmetrization. Define the effective signature rate $R_{{\rm{eff}}}:=\min\{\frac{1}{2N}, \frac{R_{QKD}}{6L}\}$, where $L$ is the length of key generated by key generation protocol in [17] and $R_{QKD}$ is the secret key rate of QKD. Note that $\frac{6L}{R_{QKD}}$ is the number of pulses required for QKD to perform symmetrization~\cite{yin2017experimental}. $R_{QKD}$ is simulated by the key rate formula in~\cite{lim2014concise}, where we choose  error-correction efficiency $f=1.22$, data post-processing block size $N=10^{10}$, secrecy $\varepsilon_{sec}=10^{-10}$ and the same experimental parameters as Fig.~\ref{fig3}.

Denote the angle between Alice-Bob and Alice-Charlie as $\theta$, the distance between Alice and Bob (Charlie) as $D_{AB}$ ($D_{AC}$), and the distance between Bob and Charlie as $D_{BC}$. In symmetric case, $D_{AB}=D_{AC}$, and  $D_{BC}=2 \sin(\frac{\theta}{2}) D_{AB}$. At short distance where $R_{QKD}$ is very high, $R_{{\rm{eff}}}$ is mainly determined by $\frac{1}{2N}$.
When $\theta$ is close to $\pi$, the transmission distance of QKD ($D_{BC}$) increases much faster than $D_{AB}$. In this case, $R_{{\rm{eff}}}$ is determined by
$\frac{R_{QKD}}{6L}$ at long distance. We simulate the case of $\theta=\frac{2}{3} \pi$ and $\theta=\pi$. As shown in Fig.~\ref{fig4}, the signature rate of \cite{PhysRevA.93.032325} is higher than that of our protocol at short distance, but when distance between the two receivers is large, the signature rate will be severely limited by the low secret key rate of QKD in the symmetrization step. By contrast, our protocol decays much slower and has a significantly longer transmission distance.

In addition, the typical experimental parameters and corresponding signature rate of some recent QDS experiments are listed in Table.~\ref{tab1}. This work shows a comparable performance with orthogonal encoding based protocol~\cite{an2019practical} even though the latter does not execute the symmetrization step. We remark that the symmetrization step is essential to the complete protocol, as demonstrated in experiments~\cite{roberts2017experimental,yin2017experimental}.

\begin{table*}
\caption{\label{tab1} { Comparison of Parameters of Recent QDS Experiments}
}
\begin{ruledtabular}
\begin{tabular}{ccccccc}
& Ref.~\cite{yin2017experimental102} & Ref.~\cite{yin2017experimental} & Ref.~\cite{roberts2017experimental} &  Ref.~\cite{ding2020280}&Ref.~\cite{an2019practical} &  This work \\ \hline
Protocol& SARG04& MDI& MDI& BB84& BB84& SARG04\\
{Repetition rate}	& 75MHz& 75MHz& 1GHz&  50MHz& 1GHz& 1GHz \\
{Transmission  distance}& 102 km& 55.6 km& 50 km &  280 km & 125 km& 125 km\\
{Security  parameter}	& $10^{-9}$ &$10^{-7}$ & $10^{-10}$ & $10^{-5}$ & $10^{-10}$ & $10^{-9}$   \\
{Signature time per bit (s)} & 66840& 149987& 45&  21407& 22.7 & 14.9  \\
{Symmetrization step} &No need & Yes& Yes& Non-execution & Non-execution & No need \\
\end{tabular}
\end{ruledtabular}
\end{table*}

\section{Conclusion}

In this paper, we have proposed a non-orthogonal encoding based  efficient quantum digital signature protocol. A novel method called post-matching is applied, which can increase the signature rate from decaying with $\eta^2$ to $\eta$.
Our protocol has a high signature rate and does not require the symmetrization operation thereby overcomes the major obstacles of existing QDS protocols. This protocol can be directly implemented with current commercially available QKD devices. Therefore, it should be the preferred solution to the application of QDS. This work is a great step for the development of quantum network with QDS. Moreover, we believe the key idea of post-matching method has the potential to be applied in various cryptographic tasks that require to establish multiparty correlations, such as multiparty quantum communication ~\cite{fu2015long}.

\section*{Funding}
National Natural Science Foundation of China (61801420); Key Research and Development Program of Guangdong Province (2020B0303040001); Fundamental Research Funds for the Central Universities.


%

\end{document}